\documentclass[]{emulateapj}

\usepackage{graphicx,times}
\slugcomment{Accepted for publication in ApJS}
\shorttitle{Convective overshoot mixing in star}
\shortauthors{Q.S. Zhang}

\begin{document}

\title{Convective overshoot mixing in stellar interior models}
\author{Q.S. Zhang \altaffilmark{1,2,3}}
\email{zqs@ynao.ac.cn(QSZ)}

\altaffiltext{1}{National Astronomical Observatories/Yunnan Observatory, Chinese Academy of Sciences, P.O. Box 110, Kunming 650011, China.}
\altaffiltext{2}{Key Laboratory for the Structure and Evolution of Celestial Objects, Chinese Academy of Sciences, Kunming, 650011, China.}
\altaffiltext{3}{Graduate University of Chinese Academy of Sciences, Beijing 100049, China.}

\begin{abstract}

The convective overshoot mixing plays an important role in stellar structure and evolution. However, the overshoot mixing is a long standing problem. The uncertainty of the overshoot mixing is one of the most uncertain factors in stellar physics. As it is well known, the convective and overshoot mixing is determined by the radial chemical component flux. In this paper, a local model of the radial chemical component flux is established based on the hydrodynamic equations and some model assumptions. The model is tested in stellar models. The main conclusions are as follows. (i) The local model shows that the convective and overshoot mixing could be regarded as a diffusion process, and the diffusion coefficient for different chemical element is the same. However, if the non-local terms, i.e., the turbulent convective transport of radial chemical component flux, are taken into account, the diffusion coefficient for each chemical element should be in general different. (ii) The diffusion coefficient of convective / overshoot mixing shows different behaviors in convection zone and in overshoot region because the characteristic length scale of the mixing is large in the convection zone and small in the overshoot region. The overshoot mixing should be regarded as a weak mixing process. (iii) The result of the diffusion coefficient of mixing is tested in stellar models. It is found that a single choice of our central mixing parameter leads to consistent results for a solar
convective envelope model as well as for core convection models of stars with mass from $2$M to $10$M.

\end{abstract}

\keywords{ convection --- turbulence --- stars: abundances --- stars: interiors }

\section{Introduction}

Many literature have shown that the convective overshoot mixing plays an important role in stellar structure and evolution. The classical treatment of the overshoot is based on the non-local mixing length theories (e.g., \citet{ss73,ma75,br81}). The convective boundary in the framework of local mixing length theory (MLT) is the location where the acceleration of the fluid elements is zero. In the framework of non-local mixing length theories, although the acceleration of fluid elements is zero ($a=0$) at the convective boundary, the velocity is not zero, thus there should be an convective overshoot region outside the convective boundary. The non-local mixing length theories trace the fluid elements overshooting from the convective unstable zone into the convective stable zone, find the location where the velocity of fluid elements is zero ($v=0$), and regard the region laying between $a=0$ and $v=0$ as the overshoot region. The convective mixing in the overshoot region, i.e., the overshoot mixing, is in usual assumed to be very efficient and leads to complete mixing. However, \citet{ren87} has argued that the theories of the mixing length type don't have enough spatial resolution to accurately describe the overshoot process, and the obtained overshoot distance in the mixing length theories is sensitive to the assumed turbulent heat flux in the overshoot region: $\nabla \approx \nabla_{ad}$ leads to extensive overshoot distance and $\nabla \approx \nabla_R$ leads to a small overshoot distance. The problem can be resolved only with self-consistent nonlocal convection theory, in which the turbulent heat flux is not external imposed, but rather determined by the theory \citep{gr93}.

Another way to describe the overshoot mixing appeared in literature is to regard the overshoot mixing as a diffusive process (e.g., \citet{fre96,sal99,la11,zha12a}). They set the diffusion coefficient of the overshoot mixing to be proportional to the local pressure scale height $H_P$ and the turbulent velocity. The turbulent velocity is set as an exponential decreasing function in the overshoot region according to numerical simulations \citep{fre96} and turbulent convection models \citep{xio85,xio89,zha12b}. Their representation of the adopted diffusion coefficient implies an assumption that the characteristic length of the overshoot mixing is on the order of $H_P$, which is same to the characteristic length of convective mixing in the unstable zone. However, this needs to be clarified.

A more reasonable way to resolve the overshoot mixing problem is to use self-consistent nonlocal convection theory in multi-component fluid. The convective / overshoot mixing is determined by the turbulent chemical component flux $\overline {{u_r}'{X_k}'}$ ($u_r$ is the radial turbulent velocity, $X_k$ is the abundance of $k$th chemical element, and $\overline {{u_r}'{X_k}'}$ is the cross-correlation function of turbulent fluctuations of $u_r$ and $X_k$). \citet{xio81} and \citet{can99,can11} have developed the self-consistent nonlocal convection theories in two-component fluid for stellar interior based on some model assumptions. The dominating equation of the turbulent chemical component flux is established in those theories. Although this way is the most reasonable one among the three to deal with the overshoot mixing, it is also the most complicated one to be applied in the calculations in stellar evolution. More advances on this way are required in order to understand and apply it.

In this paper, we present a model of the turbulent chemical component flux $\overline {{u_r}'{X_k}'}$ based on the hydrodynamic equations and some model assumptions, and discuss the properties of the overshoot and the overshoot mixing. The derivation of the model is presented in Section 2. The diffusion coefficient of the convective / overshoot mixing is presented in Section 3. The property of the convective / overshoot mixing are discussed in Section 4. The model is tested in stellar evolution in Section 5. Some discussions are presented in Section 6. The conclusions are summarized in Section 7.

\section{The equation of the convective flux of chemical composition}

Let us recall that the chemical composition through the fluid changes in two ways \citep{lau87}: (i) caused by the macroscopic motion of the fluid, which leads to purely mechanical mixing, and (ii) caused by the diffusion, i.e., the molecular / atomic transfer of the components. In this paper, we focus on the first way, i.e., the purely mechanical mixing. In the absence of molecular / atomic diffusion, the composition of a fluid element doesn't change when it moves \citep{lau87}, namely:
\begin{eqnarray}
\frac{{D{X_k}}}{{Dt}} = 0
\end{eqnarray}%
where $X_k$ is the abundance of the $k$th elements in the system (in this paper, 'k' means turbulent kinetic energy, but 'k' in subscript means the $k$th chemical element), $D/Dt$ is the Lagrange's derivation defined as follow:
\begin{eqnarray}
\frac{D}{{Dt}} \equiv \frac{\partial }{{\partial t}} + {u_i}\frac{\partial }{{\partial {x_i}}}
\end{eqnarray}%
where $u_i$ describes the velocity vector of fluid.

As it is usually used in the analysis of turbulence, a variable $A$ is split into a mean and a fluctuating part as
$A=\overline{A}+A'$. We adopt Boussinesq approximation (i.e., taking into account the the density fluctuation only in buoyancy and ignoring the compressibility of fluids), which is generally accepted in studying stellar convection theories \citep{xio97}. For this reason, this paper doesn't distinguish $\rho$ and $\overline{\rho}$, i.e., $\rho=\overline{\rho}$ in this paper. Split equation (1), note that $\overline{u_i}=0$ for the star in quasi-steady state \citep{xio89}, and use Eq.(A7), one finds:
\begin{eqnarray}
\frac{{\partial \overline {{X_k}} }}{{\partial t}} + \frac{1}{\rho }\frac{\partial }{{\partial {x_i}}}(\rho  \overline {{u_i}'{X_k}'} ) = 0
\end{eqnarray}%

In the case of spherical symmetry, equation (1) becomes:
\begin{eqnarray}
\frac{{\partial \overline {{X_k}} }}{{\partial t}} + \frac{1}{{\rho  {r^2}}}\frac{\partial }{{\partial r}}(\rho  {r^2}\overline {{u_r}'{X_k}'} ) = 0
\end{eqnarray}%

Equation (4) describes the convective mixing in stellar interior. It is clear that the convective mixing is determined by the radial chemical component flux $\overline {{u_r}'{X_k}'}$. The goal of this section is to derive the representation of the radial chemical component flux.

Let us start from the hydrodynamic equations:
\begin{eqnarray}
\frac{{\partial \rho }}{{\partial t}} + \frac{\partial }{{\partial {x_i}}}(\rho {u_i}) = 0
\end{eqnarray}%
\begin{eqnarray}
\rho \frac{{D{u_i}}}{{Dt}} =  - \frac{{\partial P}}{{\partial {x_i}}} + \rho {g_i}
 + \frac{\partial \sigma {_{ij}} }{{\partial {x_j}}}
\end{eqnarray}%
\begin{eqnarray}
\rho T\frac{{DS}}{{Dt}} = \sigma {_{ij}}\frac{{\partial {u_i}}}{{\partial {x_j}}} + \frac{\partial }{{\partial {x_j}}}(\lambda \frac{{\partial T}}{{\partial {x_j}}})
\end{eqnarray}%
where $P$ is pressure, $T$ is temperature, $S$ is entropy, $g_i$ describes the gravitational acceleration vector, $\lambda=(4acT^3)/(3\kappa\rho)$ is the thermal conduction coefficient, $\sigma {_{ij}}$ describes the viscous stress tensor defined by $\sigma {_{ij}}=\mu (\partial u_i/ \partial x_j+\partial u_j/ \partial x_i)+(\xi-2\mu/3) \delta {_{ij}} (\partial u_l/ \partial x_l)$ in which $\mu$ and $\xi$ are coefficients of viscosity \citep{lau87}.

In the energy equation, $T(DS/Dt)$ can be replaced by:
\begin{eqnarray}
  T\frac{{DS}}{{Dt}} = T{(\frac{{\partial S}}{{\partial T}})_{P,X}}\frac{{DT}}{{Dt}} + T{(\frac{{\partial S}}{{\partial P}})_{T,X}}\frac{{DP}}{{Dt}}
\\ \nonumber
  + T\sum\limits_{k'} {{{(\frac{{\partial S}}{{\partial {X_{k'}}}})}_{T,P,X - \{ {X_{k'}}\} }}\frac{{D{X_{k'}}}}{{Dt}}}
\\ \nonumber
  = {C_P}\frac{{DT}}{{Dt}} - \frac{\delta }{\rho }\frac{{DP}}{{Dt}}
\end{eqnarray}%
where $X=\{X_1,X_2,X_3,...\}$ is the chemical elements set including all independent elements (the sum of abundance of all chemical elements being unity indicates that there is one dependent element), $C_P$ is specific heat, $\delta = -(\partial ln \rho / \partial ln T)_{P,X}$ is the
expansion coefficient. Equation (1) is used in Eq.(8). The energy equation therefore can be re-written as:
\begin{eqnarray}
  \rho {C_P}\frac{{DT}}{{Dt}} = \delta \frac{{DP}}{{Dt}} + \sigma {_{ij}}\frac{{\partial {u_i}}}{{\partial {x_j}}}
  + \frac{\partial }{{\partial {x_j}}}(\lambda \frac{{\partial T}}{{\partial {x_j}}})
\end{eqnarray}%

In the Appendix, the evolutionary equation of second moment is obtained, i.e, Eq.(A10). Set $A=u_j$ and $B=X_k$ (the abundance of an independent element) in Eq.(A10), one finds the evolutionary equation of the convective chemical component flux:
\begin{eqnarray}
  \frac{{\overline D \overline {{u_j}'{X_k}'} }}{{Dt}} + \frac{1}{\rho }\frac{{\partial (\rho \overline {{u_i}'{u_j}'{X_k}'} )}}{{\partial {x_i}}} =
\\ \nonumber
  \overline {{u_j}'(\frac{{D{X_k}}}{{Dt}})'}  + \overline {{X_k}'(\frac{{D{u_j}}}{{Dt}})'}
\\ \nonumber
  - \overline {{u_i}'{u_j}'} \frac{{\partial \overline {{X_k}} }}{{\partial {x_i}}} - \overline {{u_i}'{X_k}'} \frac{{\partial \overline {{u_j}} }}{{\partial {x_i}}}
\end{eqnarray}%
where the mean Lagrange derivative is defined by $\overline{D}/Dt=(\partial / \partial t) + \overline{u_i} (\partial / \partial x_i)$. According to Eq.(1) and (6), the above equation becomes:
\begin{eqnarray}
  \frac{{\overline D \overline {{u_j}'{X_k}'} }}{{Dt}} + \frac{1}{\rho }\frac{{\partial (\rho \overline {{u_i}'{u_j}'{X_k}'} )}}{{\partial {x_i}}} =
\\ \nonumber
  \frac{1}{\rho }\overline {( - \frac{{\partial P'}}{{\partial {x_j}}} + \rho '{g_j} + \frac{\partial \sigma {_{ij}}'}{{\partial {x_i}}}){X_k}'}
\\ \nonumber
  - \overline {{u_i}'{u_j}'} \frac{{\partial \overline {{X_k}} }}{{\partial {x_i}}} - \overline {{u_i}'{X_k}'} \frac{{\partial \overline {{u_j}} }}{{\partial {x_i}}}
\end{eqnarray}%

The pressure fluctuation $P'$ describes sonic process. For the subsonic convection, the sound wave is not important for transport processes (i.e., the anelastic approximation, see \citet{sp71}), thus we ignore the pressure fluctuation, following \citet{xio81}. The density fluctuation in the buoyancy term is saved in above equation because the convection is driven by the buoyancy in stellar interior thus the buoyancy term $\rho '{g_j}$ is crucial. We use the formula as follows to model the density fluctuation in the buoyancy term:
\begin{eqnarray}
  \frac{\rho '}{\rho} =  (\frac{\partial ln \rho}{\partial ln P})_{T,X}  \frac{{P'}}{\overline{P}}   - \delta\frac{{T'}}{\overline{T}} + \sum\limits_{k'} {{\eta _{k'}}} {X_k}^\prime
 \\ \nonumber
  \approx - \delta\frac{{T'}}{\overline{T}} + \sum\limits_{k'} {{\eta _{k'}}} {X_k}^\prime;\\ \nonumber
  {\eta _{k'}} \equiv {(\frac{{\partial \ln \rho }}{{\partial {X_{k'}}}})_{T,P,X - \{ X_{k'}\} }}
   \\ \nonumber
\end{eqnarray}%
According to those, and note again $\overline{u_i}=0$, we can rewrite Eq.(11) as:
\begin{eqnarray}
  \frac{{\overline D \overline {{u_j}'{X_k}'} }}{{Dt}} + \frac{1}{\rho }\frac{\partial }{{\partial {x_i}}}(\rho \overline {{u_i}'{u_j}'{X_k}'}  - \overline {{\sigma _{ij}}'{X_k}'} ) =
\\ \nonumber
  - \frac{1}{\rho }\overline {{\sigma _{ij}}'\frac{{\partial {X_k}'}}{{\partial {x_i}}}}  - \frac{{\delta {g_j}}}{\overline{T}}\overline {T'{X_k}'}
\\ \nonumber
  + {g_j}\sum\limits_{k'} {{\eta _{k'}}} \overline {{X_{k'}}'{X_k}'}  - \overline {{u_i}'{u_j}'} \frac{{\partial \overline {{X_k}} }}{{\partial {x_i}}}
\end{eqnarray}%

Set $A=T$ and $B=X_k$ in Eq.(A10), one finds the evolutionary equation of the fluctuation $\overline{T'X_k'}$:
\begin{eqnarray}
  \frac{{\overline D \overline {T'{X_k}'} }}{{Dt}} + \frac{1}{\rho }\frac{{\partial (\rho \overline {{u_i}'T'{X_k}'} )}}{{\partial {x_i}}}
\\ \nonumber
  = \overline {T'(\frac{{D{X_k}}}{{Dt}})'}  + \overline {{X_k}'(\frac{{DT}}{{Dt}})'}
\\ \nonumber
  - \overline {T'{u_i}'} \frac{{\partial \overline {{X_k}} }}{{\partial {x_i}}} - \overline {{u_i}'{X_k}'} \frac{{\partial \overline T }}{{\partial {x_i}}}
\end{eqnarray}%

According to Eq.(1) and (7), notes that the viscous term is ignorable comparing with the dissipation term \citep{xio81,can97}, and ignores the pressure fluctuation, one finds:
\begin{eqnarray}
  \frac{{\overline D\overline {T'{X_k}^\prime } }}{{Dt}} + \frac{1}{\rho }\frac{\partial }{{\partial {x_i}}}(\rho \overline {{u_i}^\prime T'{X_k}^\prime }  - \frac{\lambda }{{{C_P}}}\overline {{X_k}^\prime \frac{{\partial T'}}{{\partial {x_j}}}} )
\\ \nonumber
  =  - \frac{\lambda }{{\rho {C_P}}}\overline {\frac{{\partial T'}}{{\partial {x_j}}}\frac{{\partial {X_k}^\prime }}{{\partial {x_j}}}}  - \frac{{\partial \overline {{X_k}} }}{{\partial {x_i}}}\overline {{u_i}^\prime T'}
\\ \nonumber
  - (\frac{{\partial \overline T }}{{\partial {x_i}}}  - \frac{\delta }{{\rho {C_P}}}\frac{{\partial \overline P }}{{\partial {x_i}}})\overline {{u_i}^\prime {X_k}^\prime }
\end{eqnarray}%

$\overline{X_kX_{k'}}$ in Eq.(13) can be modeled by analogy with \citet{can11}:
\begin{eqnarray}
  \overline {{X_k}'{X_{k'}}'}  =  - \frac{1}{2}C_1\tau( \frac{{\partial \overline {{X_{k'}}} }}{{\partial {x_i}}} \overline {{u_i}'{X_k}'} +  \frac{{\partial \overline {{X_k}} }}{{\partial {x_i}}} \overline {{u_i}'{X_{k'}}'})
\end{eqnarray}%
where $\tau=k/ \varepsilon $ is time scale of the dissipation of turbulent kinetic energy, $k$ is turbulent kinetic energy, $\varepsilon$ is the dissipation rate of turbulent kinetic energy, $C_1$ is a parameter.

The dissipation terms are modeled as follows:
\begin{eqnarray}
  \frac{\lambda }{{\rho {C_P}}}\overline {\frac{{\partial T'}}{{\partial {x_j}}}\frac{{\partial {X_k}'}}{{\partial {x_j}}}}  = {C_A}(1 + {P_e}^{ - 1}){\tau ^{ - 1}}\overline {T'{X_k}'}
\end{eqnarray}%
\begin{eqnarray}
  \frac{1}{\rho }\overline {{\sigma _{ij}}'\frac{{\partial {X_k}'}}{{\partial {x_i}}}}  = {C_B}{\tau ^{ - 1}}\overline {{u_j}'{X_k}'}
\end{eqnarray}%
where $P_e=(\rho C_P / \lambda) (k^2 / \varepsilon)$ is the P\'{e}clet number which describes the ratio of the time scale of radiative thermal conduction to the time scale of the dissipation of turbulent kinetic energy, $C_A$ and $C_B$ are parameters.

In most cases of the stellar evolutionary stage, the stellar structure can be considered as in the quasi-steady state. In the quasi-steady state, the time derivation in Eq.(13) \& (15) is zero. In additional, we assume that the turbulent fluctuations $\overline {{u_j}'X_k'}$ and $\overline {T'{X_k}'}$ are in local equilibrium. Accordingly, the equations of the structure of $\overline {{u_j}'X_k'}$ and $\overline {T'{X_k}'}$ in the quasi-steady local equilibrium (local model) are as follows:
\begin{eqnarray}
  \overline {{u_r}'T'} \frac{{\partial \overline {{X_k}} }}{{\partial r}} - \frac{{\overline T }}{{{H_P}}}(\nabla  - {\nabla _{ad}})\overline {{u_r}'{X_k}'}
\\ \nonumber
  + {C_A}(1 + {P_e}^{ - 1}){\tau ^{ - 1}}\overline {T'{X_k}'}=0
\end{eqnarray}%
\begin{eqnarray}
  \overline {{u_r}'{u_r}'} \frac{{\partial \overline {{X_k}} }}{{\partial r}} -\frac{{\delta g}}{\overline{T}}\overline {T'{X_k}'}
\\ \nonumber
  + ({C_B}{\tau ^{ - 1}} + \frac{1}{2}\frac{g}{{{H_P}}}{C_1}\phi \tau )\overline {{u_r}'{X_k}'}
\\ \nonumber
  - \frac{1}{2}g{C_1}\tau \sum\limits_{k'} {\eta _{k'}\overline {{u_r}'{X_{k'}}'} } \frac{{\partial \overline {{X_k}} }}{{\partial r}}=0
\end{eqnarray}%
where $\phi  = \sum\limits_{k'} {{\eta _{k'}}(\partial \overline {{X_{k'}}} / \partial \ln P)} $, $g=-g_r=GM_r/r^2$ is the gravitational acceleration, $\nabla = dlnT/dlnP$ is the real temperature gradient in stellar interior, and $\nabla _{ad}=( \partial lnT / \partial lnP)_S$ is the adiabatic temperature gradient.

\section{The diffusion coefficient of convection induced mixing}

It is not difficult to obtain the the solution of the convective flux in Eq.(19) \& (20) as follows:
\begin{eqnarray}
  \overline {{u_r}'{X_k}'}  = -{D_k}\frac{{\partial \overline {{X_k}} }}{{\partial r}}
\end{eqnarray}%
where
\begin{eqnarray}
  {D_k} = [{C_A}(1 + {P_e}^{ - 1}){\tau ^{ - 1}}\overline {{u_r}'{u_r}'}  + \frac{{\delta g}}{{\overline T }}\overline {{u_r}'T'}
\\ \nonumber
  - \frac{1}{2}{C_1}{C_A}(1 + {P_e}^{ - 1})g\sum\limits_{k'} {({\eta _{k'}}\overline {{u_r}'{X_{k'}}'} )}]
\\ \nonumber
  [{C_B}{C_A}(1 + {P_e}^{ - 1}){\tau ^{ - 2}}
  + \frac{1}{2}\frac{g}{{{H_P}}}{C_1}{C_A}(1 + {P_e}^{ - 1})\phi
\\ \nonumber
  - \frac{{\delta g}}{{{H_P}}}(\nabla  - {\nabla _{ad}})]^{-1}
\end{eqnarray}%
The value of $D_k$ still can't be work out directly because the turbulent variable $\overline {{u_r}'{X_{k'}}'}$ is present in the right hand side. However, the right hand side is independent of 'k' in the representation of $D_k$. This indicates that $D_k$ for all chemical element are the same in the local model, i.e., $D_k=D$. Accordingly, one finds $\overline {{u_r}'{X_{k'}}'} ({\partial \overline {{X_k}} }/{\partial r})=\overline {{u_r}'{X_{k}}'} ({\partial \overline {{X_{k'}}} }/{\partial r})$ in Eq.(20) and then gets the representation of $D$:
\begin{eqnarray}
  D = [{C_A}(1 + {P_e}^{ - 1}){\tau ^{ - 1}}\overline {{u_r}'{u_r}'}  + \frac{{\delta g}}{{\overline T }}\overline {{u_r}'T'}]
\\ \nonumber
  \{{C_B}{C_A}(1 + {P_e}^{ - 1}){\tau ^{ - 2}}
\\ \nonumber
  - \frac{{\delta g}}{{{H_P}}}[\nabla  - {\nabla _{ad}} - \frac{{{C_1}{C_A}}}{\delta }(1 + {P_e}^{ - 1})\phi ]\}^{-1}
\end{eqnarray}%
This result indicates that the convective / overshoot mixing can be regarded as a diffusion process in the local model.

The representation of Eq.(23) hints us to set $C_1=1/C_A$. Note that the region with $P_e \ll 1$ is in the stellar surface convection zone and thus there is almost $\phi = 0$, one finds the representation of the diffusion coefficient as follow:
\begin{eqnarray}
  D = \frac{{ 2\omega{C_A}(1 + {P_e}^{ - 1}) + \frac{{\delta g}}{\overline{T}}\frac{\overline {{u_r}'T'}}{\varepsilon} }}{{ {C_A}{C_B}(1 + {P_e}^{ - 1})+ N^2\tau ^2}} \frac{k^2}{\varepsilon}
\end{eqnarray}%
where $\omega=\overline {{u_r}'{u_r}'}/(2k)$ is the anisotropic degree, $N^2=- \delta g(\nabla  - {\nabla _{ad}} - \phi /\delta )/{H_P}$ describes Brunt-V\"{a}is\"{a}l\"{a} frequency. The turbulent properties (i.e, the turbulent heat flux $\overline {{u_r}'T'}$, the turbulent kinetic energy $k$ and the turbulent dissipation rate $\varepsilon$) are required for calculating $D$. Those variables can be determined by the turbulent convection models (TCMs) (e.g., \citet{xio85,can97,den06,li07,li12}). Here, we adopt Li \& Yang's (2007) TCM. Set the model of $\overline {{u_r}'T'}$ and $\overline {T'T'}$ to be local, i.e., $C_{t1}=0$ and $C_{e1}=0$ in the TCM, one obtains $\overline {{u_r}'T'}$ as follow:
\begin{eqnarray}
  \overline {{u_r}'T'}  =  - \frac{{2\omega {C_e}(1 + {P_e}^{ - 1})}}{{{C_e}{C_t}{{(1 + {P_e}^{ - 1})}^2} + N^2{\tau ^2}}}k\tau
\\ \nonumber
  [ - \frac{T}{{{H_P}}}(\nabla  - {\nabla _{ad}})]
\end{eqnarray}%
where $C_e$, $C_t$, $C_{e1}$ and $C_{t1}$ are model parameters in Li \& Yang's (2007) TCM (see Eq.(6), (7), (10) \& (11) in their paper).

Therefore, the diffusion coefficient of the convective / overshoot mixing $D$ is as follow:
\begin{eqnarray}
  D = 2\omega (1 + {P_e}^{ - 1})\frac{{{C_A} - \frac{{{C_e}{N^2}{\tau ^2}}}{{{C_e}{C_t}{{(1 + {P_e}^{ - 1})}^2} + {N^2}{\tau ^2}}}}}{{{C_B}{C_A}(1 + {P_e}^{ - 1}) + {N^2}{\tau ^2}}}(\tau k)
\end{eqnarray}%

In stellar interior, the most important case for the convective / overshoot mixing is in the region with $P_e \gg 1$. Now, let us focus on the case of $P_e \gg 1$. In the convection zone with $P_e \gg 1$, Li \& Yang's (2007) TCM shows ${N^2}{\tau ^2} \approx  - {C_e}{C_t}/(2{C_e}\omega  + 1)$ which can be worked out via simple algebra using the information in Appendix A in \citet{zha12b}. In the overshoot region, TCMs show that the temperature gradient is close to the radiative temperature gradient \citep{xio89,xio01,den06,zha12b}, which is consistent with helioseismic investigation \citep{chr11}, thus $N^2 \tau ^2 \gg 1$. Accordingly, a simple representation of $D$ is as follows:
\begin{eqnarray}
  D \approx \frac{{2\omega {C_A} + 1}}{{{C_B}{C_A} - \frac{{{C_e}{C_t}}}{{2{C_e}\omega  + 1}}}}\frac{{{k^2}}}{\varepsilon }
\end{eqnarray}%
in the convection zone, and:
\begin{eqnarray}
  D \approx2 \omega ({C_A} - {C_e})\frac{\varepsilon }{{{N^2}}}
\end{eqnarray}%
in the overshoot region.

\section{The characteristic time scale and length scale of convective and overshoot mixing}

In the above section, it is found that the convective mixing can be deal with a diffusion process and the diffusion coefficient, i.e., Eq.(26), shows different behaviors in convection zone and overshoot region. In this section, we discuss the physical meaning of the diffusion coefficient and the mechanism of convective and overshoot mixing.

In order to understand the diffusion mixing process, one should study the characteristic time scale $\tau_{Mix}$ and the characteristic length of mixing $L_{Mix}$. The convection induced mixing is 'purely mechanic mixing', thus it is reasonable to regard the mixing as a consequence of the turbulent dissipation. Therefore, the characteristic time scale of mixing should be comparable with the time scale of turbulent dissipation of kinetic energy, i.e., $\tau_{Mix} \sim \tau$.
The characteristic length of the mixing should be the length of the radial range of the movement of fluid element, because the fluid element is finally dissolved in this range.

In the convection zone, the buoyancy impulses the radial movements of fluid elements, thus the radial length of the movement of an fluid element is the typical velocity multiplies the its lifetime which is comparable with the turbulent dissipation time scale of kinetic energy, i.e., $L_{Mix} \sim \sqrt{k}\tau$. Therefore, the diffusion coefficient should be: $D \propto L_{Mix}^2/ \tau_{Mix} \sim k^2/\varepsilon$.

In the overshoot region, fluid elements gain kinetic energy via turbulent diffusion of kinetic energy, and move (in radial direction) around their equilibrium location, because the buoyancy prevents the radial movements of fluid elements. In this scene, the length of the radial range of the movement, also the characteristic length of the mixing $L_{Mix}$, can be estimated as $L_{Mix} \sim \sqrt{k}/N$. Accordingly, the diffusion coefficient should be: $D \propto L_{Mix}^2/ \tau_{Mix} \sim \varepsilon/N^2$.

It is necessary to discuss the relation between the turbulent heat transport and the matter mixing.
In the high $P_e$ region, the time scale of radiative thermal conduction is much longer than the time scale of the dissipation of turbulent kinetic energy, thus there is almost no radiative heat exchange between fluid elements and the surrounding medium. Therefore, the heat exchange is a consequence of the matter mixing, or further, the turbulent dissipation. When the turbulent kinetic energy of a fluid element is completely dissipated, the fluid element is dissolved into the surrounding medium, thus it contributes its chemical composition and also its entropy to surrounding medium. The former causes the matter mixing, and the latter, i.e., entropy mixing, causes the heat exchange which is described by the turbulent heat transport. The turbulent mixing is not only on matter, but also on entropy. The turbulent heat transport can be regarded as a consequence of the matter mixing. When the mixing is effective, the entropy mixing leads to isentropic region, thus the temperature gradient should be close to the adiabatic temperature gradient, and the chemical composition should be almost uniformed. On the contrary, when the mixing is ineffective, the temperature gradient should be close to the radiative temperature gradient, and the chemical composition in the region, which is not chemically uniformed before taking into account the turbulent mixing, should be still not chemically uniformed. According to the discussions, the diffusion coefficient for heat transport should be similar to the case of matter mixing. Equation (25) shows that the diffusion coefficient $D_T$ for the convective / overshoot heat transport is as follow:
\begin{eqnarray}
  D_T  =  \frac{{2\omega {C_e}(1 + {P_e}^{ - 1})}}{{{C_e}{C_t}{{(1 + {P_e}^{ - 1})}^2} + N^2{\tau ^2}}}(k\tau)
\end{eqnarray}%
The representation of $D_T$ also shows that $D_T \propto k^2/\varepsilon$ in the convection zone and $D_T \propto \varepsilon/N^2$ in the overshoot region, as similar as the case of matter mixing.

Those phenomenological discussions explain the representation of Eqs.(27-28). It is indicated in Eq.(26) that the convective mixing in the convection zone ($|N^2 \tau ^2| \sim 1$) is much more efficient than the situation in the overshoot region  ($N^2 \tau ^2\gg 1 $). This is because of the different characteristic length of mixing. In the overshoot region, the radial movements of fluid elements are around their equilibrium location, thus the characteristic length of mixing is much shorter than the situation in the convection zone. This could be the reason of why an excessively small parameter $C_X=10^{-10}$ is adopted in order to fit some observations when $D=C_X H_P \sqrt{k}$  ($\approx C_X k^2/\varepsilon$, similar to Eq.(27)) is applied to the overshoot mixing (see \citet{zha12a,zqs12a,zqs12b}).

\section{Applications in stellar models}

Usually, the turbulent kinetic energy $k$ in the convection zone is on the magnitude order $10^6$ or more, and the dissipation rate $\varepsilon=k^{3/2}/(\alpha H_P)$, where $H_P$ is around the magnitude order $10^9$ and $\alpha \sim 1$. These lead to that the diffusion coefficient $D$ is on the magnitude order $10^{12}$ or more in the convection zone. The size of convection zone is about $L\sim(10^{-1}\sim10^1) H_P$. Therefore, the time scale of complete mixing in convection zone is about $t_{Mix}\sim L^2/D\sim (10^4\sim10^8)$ seconds or less, which is short enough in most cases in stellar evolution, leading to the complete mixed convection zone. The most interesting effect of the diffusion coefficient Eq.(26) is on the overshoot region. Equation (28) shows that the diffusion coefficient of overshoot mixing is proportional to $(C_A-C_e)$, thus we define a new parameter:
\begin{eqnarray}
  C_{X1}\equiv C_A-C_e
\end{eqnarray}%

In this section, we try to find the proper value of $C_{X1}$. The diffusive overshoot mixing based on the diffusion coefficient Eq.(26) is tested in two cases: the solar model and the Hertzsprung-Russell (H-R) diagram of main sequence stars with convective core. The first corresponds to the convective envelope downward overshoot mixing, and the second one corresponds to the convective core overshoot mixing. Recently, \citet{zha12a} and \citet{zqs12a,zqs12b} have studied the overshoot mixing in stellar model of low-mass stars by using the diffusion coefficient described by $D=C_X H_P \sqrt{k}$ and found that the parameter $C_X$ should be on the magnitude order $10^{-10}$. The excessively small parameter indicates that the representation of adopted diffusion coefficient, i.e., $D=C_X H_P \sqrt{k}$, seems to be not physically suitable.

Li \& Yang's (2007) TCM is adopted in the stellar modeling. \citet{paz69} evolution code modified by \citet{zqs12b} is used to model stars. The modified code can solve the combination of stellar structure equations and the TCM in stellar evolution models. The diffusion coefficient of the overshoot mixing is describe by Eq.(26) in the overshoot region. Parameters of the TCM are same to \citet{zqs12b}, except for $\alpha$. The value of $\alpha$ is $0.878$ resulting from the solar calibration based on the TCM and the diffusion coefficient Eq.(26). The model parameter $C_B$ is insensitive to the diffusion coefficient in the overshoot region. It is set to be $C_B=2 C_t$.

The OPAL equation of state \citep{rog96}, the OPAL opacity tables for high temperatures \citep{igl96},
and the Alexander's opacity tables for low temperatures \citep{ale94} are used. The composition mixture is assumed to be the same as the solar mixture \citep{gs98}. The elements settling \citep{tho94} is used in the solar models.

\subsection{The convective envelope overshoot: the solar model}

For the solar model, there are two properties might be connected with the overshoot mixing below the base of the convective envelope, i.e., the sound speed profile and the surface Li abundance (see e.g., \citet{sch99,br99,zha12a}). There are some different proposal to solve the Li problem. However, the recent helioseismic investigation (i.e., \citet{chr11}) indicates that the convective overshoot mixing should be taken into account when study the Li abundance \citep{zqs12a}. The solar models evolve from ZAMS to the present solar age $4.6Gyr$. Three evolutionary series of the solar models are calculated, include the standard solar model (MLT is used, no overshoot) and two solar models based on the TCM and overshoot mixing with $C_{X1}=10^{-3}$ and $C_{X1}=10^{-2}$. The radius, luminosity and the boundary of the convection zone of the present solar models are calibrated to fit the observations ($R_{\odot}=6.96\times10^{10}$, $L_{\odot}=3.846\times10^{33}$) and helioseismic inversion (the base of the solar convection zone $R_{CZ}=0.7135R_{\odot}$, \citet{chr91,ba97,ba98}) via the iterations of initial hydrogen abundance, metallicity, and the TCM parameter $\alpha$,.

The overshoot mixing solar model with $C_{X1}=10^{-3}$ shows $0.5$ dex Li depletion in the main sequence stage, and the model with $C_{X1}=10^{-2}$ shows $1.4$ dex. It has been found that the pre-main sequence Li depletion is about $1$ dex \citep{dm94,pt02,zha12a}. But the pre-main sequence Li depletion is sensitive to the opacity, there is an uncertainty about $0.5$ dex \citep{dm94}. The observations \citep{an89,ags09} require about $2$ dex Li depletion. Accordingly, the main sequence Li depletion should be $0.5\sim1.5$ dex. This indicates that the parameter $C_{X1}$ should be in the range of $10^{-3}\sim10^{-2}$. The differences between the sound speed of the solar models and the helioseismic inversion are shown in Fig.1. It can be found in Fig.1 that the sound speed has been improved in both two TCM solar models with overshoot mixing. The differences of the sound speed of model below the base of the convection zone are reduced to be less than $0.2\%$ in both of the overshoot solar models.

\begin{figure}
\includegraphics[scale=0.7]{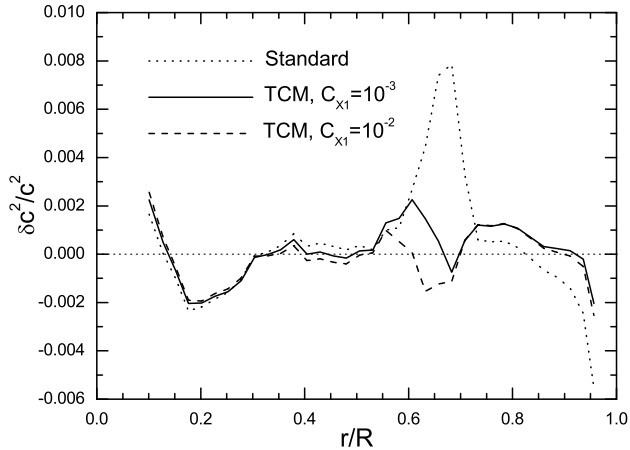}
\caption{The sound speed differences in the solar models. $\delta c^2/c^2=(c_\odot^2 - c_{model}^2)/c_\odot^2$ is the sound speed difference. $c_{model}$ is the sound speed of the model. $c_\odot$ is the helioseismic inversion of sound speed, taken from \citet{ba09}.}
\end{figure}

\subsection{The convective core overshoot: evolutionary tracks of main sequence stars in the H-R diagram}

For the stars with convective core, it is well known that the core overshoot mixing leads to significant effects on the stellar evolution. The classical treatment on the overshoot is to set an overshoot region outside the convective core and the the overshoot region is assumed to be complete mixed as well as the convective core. The length of the overshoot region is usually set as $l_{OV} = \alpha_{OV}H_P$, where $\alpha_{OV}$ is a parameter. The proper range of the value of $\alpha_{OV}$ can be derived by comparing the stellar models with corresponding observation data. \citet{st91} has found an upper limit of $0.4H_P$ for the core overshoot region (for the classically complete overshoot mixing). \citet{cl07} has investigated the length of the classical core overshoot region by using the observations of double-lined eclipsing binaries, and the results also indicated that, when the error bars are taken into account, the core overshoot region based on the complete mixing seems to be less than $0.4H_P$. Claret's (2007) results and some of Stothers's (1991) are based on the properties of stellar evolutionary tracks in the H-R diagram. Therefore, their results require that the satisfied stellar evolutionary track of the star with a convective core should locate between the evolutionary track of $0.4H_P$ completely overshoot mixing stellar model and the evolutionary track of standard stellar model without overshoot.

\begin{figure}
\includegraphics[scale=0.6]{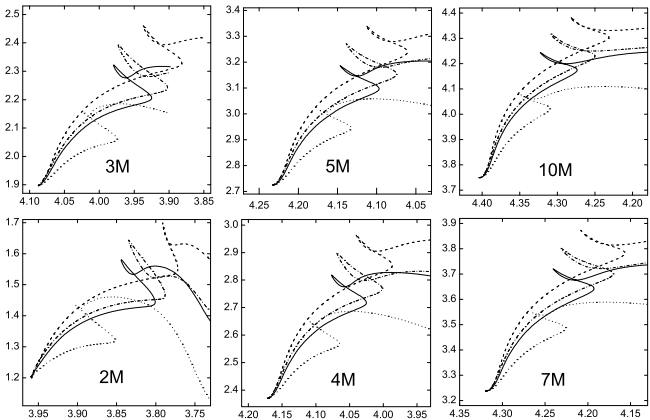}
\caption{Stellar evolutionary tracks in the HR diagram of in the main sequence stage. The dotted lines correspond to the standard stellar models, in which the convection heat flux is calculated by using the MLT and the overshoot mixing is absent. The dot-dashed lines correspond to the classical overshoot stellar models, in which the convection heat flux is calculated by using the MLT and the overshoot leads to complete mixing in $0.4H_P$ outside the boundary of the convective core. The solid and dashed lines correspond to the stellar model with the TCM and diffusive overshoot mixing. The value of $C_{X1}$ for solid and dashed lines is $10^{-3}$ and $10^{-2}$, respectively.  }
\end{figure}

Figure 2 shows the evolutionary tracks of $2M$, $3M$, $4M$, $5M$, $7M$, and $10M$ stars. Stellar models in four cases are calculated: the standard stellar models (MLT is used, no overshoot), the classical overshoot stellar models (MLT is used, completely overshoot mixing in $0.4H_P$ outside the Schwarzchild convective boundary), the TCM stellar models with the diffusive overshoot mixing with $C_{X1}=10^{-3}$ and $C_{X1}=10^{-2}$. The initial chemical composition is hydrogen abundance $X_H=0.7$ and metallicity $Z=0.02$ for all stellar models. The dotted lines correspond to the standard stellar models. The dot-dashed lines correspond to the classical overshoot stellar models. The solid and dashed lines correspond to the TCM stellar model with the diffusive overshoot mixing, where $C_{X1}=10^{-3}$ for solid and $C_{X1}=10^{-2}$ for dashed lines respectively. It can be found that only the the TCM stellar models with $C_{X1}=10^{-3}$ locate between the evolutionary tracks of $0.4H_P$ completely overshoot mixing stellar models and the evolutionary tracks of standard stellar model. All of the TCM stellar model with $C_{X1}=10^{-2}$ show too strong efficiency of overshoot mixing. Therefore, based on Li \& Yang's (2007) TCM and adopted TCM parameters, the proper value of $C_{X1}$ should be about $10^{-3}$ or less.

\section{Discussions}

\subsection{On the overshoot}

The 'ballistic' overshoot models (e.g., \citet{ss73,ma75,br81}) investigate how far a fluid element from the convective unstable zone can go into the stable zone, and define the location where the velocity of the fluid element is zero as the boundary of the overshoot region. In this description, the turbulent convective transport of turbulent kinetic energy is not taken into account. The overshoot flows stir the surrounding medium and transport turbulent momentum and kinetic energy, thus the medium originally locating in the stable zone should move. And, the moving medium go on stirring and transporting turbulent momentum and kinetic energy. According to this mechanism, the real overshoot region should be larger than the prediction by 'ballistic' overshoot models. This is consistent with turbulent convection models, which show very extensive overshoot region that is much more extensive than the prediction by 'ballistic' overshoot models. In this sense, one should regard the 'overshoot region' as the overshoot of turbulent kinetic energy rather than the overshoot of the fluid elements from the convective unstable zone.

According to above discussions, the flows in the overshoot region can be classified as two kinds: (i) the flow overshooting from the convective unstable zone ,and (ii) the flow caused by the turbulent convective transport of turbulent kinetic energy. The 'ballistic' overshoot models focus only on the first type of flow. Fluid elements in the second type of flow are originally staying and locating in the stable zone. They gain turbulent kinetic energy via the turbulent convective transport. Because the buoyancy in the stable region prevents the radial movement of fluid elements, they should move (in the radial direction) around their equilibrium location. The radial range of their movements is much shorter than that in the convective unstable zone, thus the effect of the mixing (heat and matter transport) is much lower than that in the convective unstable zone. This indicates that, in the overshoot region, the temperature gradient is more closer to the radiative temperature gradient than to the adiabatic one, and the chemical composition may not be uniform. The fluid element moves around its equilibrium location, thus the radial velocity is the maximum when the fluid element is in its equilibrium location where the turbulent temperature fluctuation is zero, and the radial velocity is zero when the turbulent temperature fluctuation is the maximum. This indicates that the correlation coefficient between turbulent radial velocity and turbulent temperature fluctuation is much smaller than that in the convective unstable zone. It is for the reason that, in the convective unstable zone, the fluid element is accelerated by the buoyancy and then turbulent temperature fluctuation and turbulent radial velocity increase in the same time, leading to significant correlativity between turbulent temperature fluctuation and turbulent radial velocity in the convective unstable zone.

The low efficiency of the mixing (heat and matter transport) leads to $\nabla \approx \nabla_{R}$ and weak matter mixing in the overshoot region. The former is consistent with turbulent convection models, which show $\nabla \approx \nabla_{R}$ in the overshoot region with $P_e \gg 1$ (e.g., \citet{xio85,xio01,zcg12,zqs12b}). And, the latter is favored by the helioseismic results: the helioseismic investigation \citep{chr11} shows $0.37H_P$ overshoot below the base of the solar convection zone, and \citet{zqs12a} has found that complete mixing in the overshoot region with the length $0.37H_P$ leads to excessive Li depletion in solar model. The low correlation coefficient between turbulent radial velocity and turbulent temperature fluctuation in the overshoot region is consistent with numerical simulations (e.g., \citet{sin95,me07}) and turbulent convection models (e.g., \citet{xio85,xio89,xio92,xio01,den06,zha12b}).

\subsection{On the parameter $C_{X1}$}

The results of Section 5 show that, based on Li \& Yang's (2007) TCM and adopted TCM parameters, $C_{X1}\equiv{C_A} - {C_e}$ seems to be about $\sim 10^{-3}$ in order to satisfy some observational restrictions. The value of $C_{X1}$ being small indicates that $C_A \approx C_e$. This is understandable. Let us recall the definition of $C_e$, i.e., Eq.(7) in \citet{li07}:
\begin{eqnarray}
  \frac{\lambda }{{\rho {C_P}}}\overline {\frac{{\partial T'}}{{\partial {x_j}}}\frac{{\partial T'}}{{\partial {x_j}}}}  = {C_e}(1 + {P_e}^{ - 1}){\tau ^{ - 1}}\overline {T'T'}
\end{eqnarray}%
This is similar with the definition of $C_A$, i.e., Eq.(17). It should be noticed that $\triangle T $ (the difference of the temperature between a fluid element and the surrounding medium) is strongly correlated with $\triangle X_k $ (the difference of the abundance of $k$th chemical element between the fluid element and the surrounding medium), since both of them are approximately proportional to the radial distance between the location of the fluid element and its equilibrium location. This indicates that the turbulent temperature fluctuation is strongly correlated with the turbulent abundance fluctuation, which means that $T'$ is approximately proportional to $X_k'$. Taking the strong correlation between $T'$ and $X_k'$ into Eq.(31), comparing with Eq.(17), one finds $C_A \approx C_e$.

\subsection{On the nonlocal model}

The diffusion coefficient in Section 3 is based on Eq.(19-20), which are the local limit of Eq.(13) \& (15). An conclusion based on the local model is that the diffusion coefficient for different chemical element is the same, namely, $D_k=D$. When the turbulent convective transport terms, i.e., $(\partial \rho\overline {{u_i}'T'{X_k}'}/\partial x_i)$ and $(\partial \rho\overline {{u_i}'{u_j}'{X_k}'}/\partial x_i)$ in Eq.(13) \& (15), are taken into account, this conclusion doesn't hold in general condition. This is because the turbulent convective transport terms depend on high order radial derivations of chemical abundance. Therefore, in general, the efficient of convective and overshoot mixing is different for each chemical element. However, there is a special case, i.e., when the abundance of $i$th chemical element $X_i$ plus the abundance of $j$th chemical element $X_j$ is a constant in the stellar interior. $X_i$ and $X_j$ satisfy $\partial X_i/\partial r = - \partial X_j/\partial r$ because $X_i+X_j$ is constant. In this case, $D_i = D_j$ always holds whether the turbulent convective transport terms are taken into account or not. This is because the equation:
\begin{eqnarray}
  0=\overline {{u_r}'(X_i+X_j)'}=\overline {{u_r}'{X_i}'}+\overline {{u_r}'{X_j}'}
\\ \nonumber
  =-D_i \frac{\partial X_i}{\partial r}-D_j \frac{\partial X_j}{\partial r}
  =-(D_i-D_j) \frac{\partial X_i}{\partial r}
\end{eqnarray}%
mathematically requires $D_i = D_j$. An example is the calculation of standard stellar model of main sequence star without the settling and atomic diffusion, in which the metallicity is uniform in the stellar interior and then the sum of the hydrogen abundance and the helium abundance is constant.

\section{Conclusions}

In this paper, we studied the convective overshoot mixing in stellar interior. The convective and overshoot mixing is determined by the radial chemical component flux $\overline {{u_r}'{X_k}'}$. A local model of the radial chemical component flux is obtained based on the hydrodynamic equations and some model assumptions. The model is tested in stellar models. The main conclusions are as follows:

1. The local model of the radial chemical component flux shows that the convective and overshoot mixing could be regarded as a diffusion process, and the diffusion coefficient for different chemical element is the same. However, if the non-local terms, i.e., $(\partial \rho\overline {{u_i}'T'{X_k}'}/\partial x_i)$ and $(\partial \rho\overline {{u_i}'{u_j}'{X_k}'}/\partial x_i)$, are taken into account, the diffusion coefficient for each chemical element should be in general different.

2. The diffusion coefficient of convective / overshoot mixing shows different behaviors in convection zone and in overshoot region. In the convection zone, the characteristic length scale of the mixing is large, thus the diffusion coefficient is high enough to ensure that the convection zone is completely mixed in most cases. However, in the overshoot region, the characteristic length scale of the mixing is much shorter than in the case of convection zone, thus the diffusion coefficient is much lower. The overshoot mixing should be regarded as a weak mixing process.

3. The representation of the diffusion coefficient of mixing is tested in stellar models. It is found that, based on Li \& Yang's (2007) TCM and adopted TCM parameters, the key parameter of overshoot mixing $C_{X1}\equiv C_A-C_e$ should be about $10^{-3}$ in order to satisfy some observational restrictions. The small value of $C_{X1}$ means $C_A \approx C_e$, which is thought to be reasonable according to the properties of the convection motion in stellar interior.

\acknowledgments

Q. S. Z. thanks the anonymous referee for careful reading of the manuscript and comments which improved the original version.
This work is co-sponsored by the National Natural Science Foundation of China through grant No.10973035 and Science Foundation of Yunnan Observatory No.Y0ZX011009 and No.Y1ZX011007. Q .S. Z. is funded by the exchange program between Chinese Academy of Sciences and the Danish Rectors¡¯ Conference (Universities Denmark). Fruitful discussions with Y. Li are highly appreciated.

\appendix

\section{Some properties of the fluctuations}

Some basic properties:
\begin{eqnarray}
  A=\overline{A}+A'
\end{eqnarray}%
\begin{eqnarray}
  \overline{AB}=\overline{A}\cdot\overline{B}+\overline{A'B'}
\end{eqnarray}%
\begin{eqnarray}
  \overline{A'}=\overline{A-\overline{A}}=\overline{A}-\overline{A}=0
\end{eqnarray}%
\begin{eqnarray}
  \overline{A'B'}=\overline{A'B'}+0=\overline{A'B'}+\overline{A'\overline{B}}=\overline{A'B}
\end{eqnarray}%
\begin{eqnarray}
  (AB)'=(\overline{A}+A')(\overline{B}+B')-\overline{A}\cdot\overline{B}-\overline{A'B'}
\\ \nonumber
  =A'\overline{B}+B'\overline{A}+(A'B')'
\end{eqnarray}%

According to the continuity equation and Boussinesq approximation (ignoring the density fluctuation), one finds:
\begin{eqnarray}
  \frac{\partial (\rho {u_i})'}{{\partial {x_i}}}=0
\end{eqnarray}%

 As a consequence:
\begin{eqnarray}
  {u_i}'\frac{{\partial A}}{{\partial {x_i}}} = \frac{1}{\rho }\frac{{\partial (\rho {u_i}'A)}}{{\partial {x_i}}}
\end{eqnarray}%

The mean Lagrange's derivation:
\begin{eqnarray}
  \frac{{\overline D A}}{{Dt}} \equiv \frac{{\partial A}}{{\partial t}} + \overline {{u_i}} \frac{{\partial A}}{{\partial {x_i}}}  = \frac{{DA}}{{Dt}} - {u_i}'\frac{{\partial A}}{{\partial {x_i}}}
\\ \nonumber
  = \frac{{DA}}{{Dt}} - \frac{1}{\rho }\frac{{\partial (\rho {u_i}'A)}}{{\partial {x_i}}}
\end{eqnarray}%

The evolutionary equation of the second-moment:
\begin{eqnarray}
  \frac{{\overline D (\overline {A'B'}) }}{{Dt}} = \overline {\frac{{\overline D (A'B')}}{{Dt}}}
\\ \nonumber
  = \overline {A'\frac{{\overline D B'}}{{Dt}} + B'\frac{{\overline D A'}}{{Dt}}}
  = \overline {A'\frac{{\overline D B}}{{Dt}} + B'\frac{{\overline D A}}{{Dt}}}
\\ \nonumber
  = \overline {A'\frac{{DB}}{{Dt}}}  - \overline {A'{u_i}'\frac{{\partial B}}{{\partial {x_i}}}}  + \overline {B'\frac{{DA}}{{Dt}}}
  - \overline {B'{u_i}'\frac{{\partial A}}{{\partial {x_i}}}}
\\ \nonumber
\\ \nonumber
  = \overline {A'(\frac{{DB}}{{Dt}})'}  - (\overline {A'{u_i}'\frac{{\partial B'}}{{\partial {x_i}}}}
  + \overline {A'{u_i}'} \frac{{\partial \overline B }}{{\partial {x_i}}})
\\ \nonumber
 + \overline {B'(\frac{{DA}}{{Dt}})'}  - (\overline {B'{u_i}'\frac{{\partial A'}}{{\partial {x_i}}}}  + \overline {B'{u_i}'} \frac{{\partial \overline A }}{{\partial {x_i}}})
\\ \nonumber
  = \overline {A'(\frac{{DB}}{{Dt}})'}  + \overline {B'(\frac{{DA}}{{Dt}})'}
\\ \nonumber
  - \overline {A'{u_i}'} \frac{{\partial \overline B }}{{\partial {x_i}}} - \overline {B'{u_i}'} \frac{{\partial \overline A }}{{\partial {x_i}}} - \frac{1}{\rho }\frac{{\partial (\rho \overline {{u_i}'A'B'} )}}{{\partial {x_i}}}
\end{eqnarray}%

In the derivation of the above equation, Eqs.(A4),(A7) \& (A8) are used. The final result is as follow:
\begin{eqnarray}
  \frac{{\overline D (\overline {A'B'}) }}{{Dt}} + \frac{1}{\rho }\frac{{\partial (\rho \overline {{u_i}'A'B'} )}}{{\partial {x_i}}} =
\\ \nonumber
  \overline {A'(\frac{{DB}}{{Dt}})'}  + \overline {B'(\frac{{DA}}{{Dt}})'}  - \overline {A'{u_i}'} \frac{{\partial \overline B }}{{\partial {x_i}}} - \overline {B'{u_i}'} \frac{{\partial \overline A }}{{\partial {x_i}}}
\end{eqnarray}%

The above equation describes the evolution of the second moment $\overline {A'B'}$. The second term in the left hand side is the turbulent convective transport, the first two terms in the right hand side are the turbulent local productive terms, the last two terms are the contributions from the mean field.

\end{document}